\documentclass[final]{svjour2}                    
\smartqed  
\usepackage{graphicx}
\usepackage{sidecap} 
\usepackage{caption}

\begin{document}

\title{Bootstrapping topology and systemic risk of complex network using the fitness model}

%


\author{Nicol\`o Musmeci \and Stefano Battiston \and Guido Caldarelli \and Michelangelo Puliga 
\and Andrea Gabrielli}


\institute{N. Musmeci \at
              Dipartimento di Fisica Universit\`a ``Sapienza'' P.le A. Moro 5, 00185 Rome, Italy. 
              \email{nik.musmeci87@gmail.com}           
           \and
           S. Battiston \at
              Chair of Systems Design, ETH Zurich, Weinbergstrasse 56/58, 8092, Zurich, Switzerland. 
              \email{sbattiston@ethz.ch}           
           \and
           G. Caldarelli \at
              IMT Alti Studi Lucca, Piazza S. Ponziano 6 55100, Lucca, Italy. 
              ISC-CNR UdR Sapienza, P.le A. Moro 5, 00185 Rome, Italy. 
              London Institute for Mathematical Sciences 35a South Street Mayfair London  W1K 2XF, UK. 
              \email{guido.caldarelli@imtlucca.it}           
	    \and
           M. Puliga \at
              Chair of Systems Design, ETH Zurich, Weinbergstrasse 56/58, 8092, Zurich, Switzerland. 
              \email{puligam@ethz.ch}           
              \and
           A. Gabrielli \at
              ISC-CNR UdR Sapienza, P.le A. Moro 5, 00185 Rome, Italy. 
	      IMT Alti Studi Lucca, Piazza S. Ponziano 6 55100, Lucca, Italy. 
	      London Institute for Mathematical Sciences 35a South Street Mayfair London  W1K 2XF, UK. 
              \email{andrea.gabrielli@roma1.infn.it}           
}

\date{Received: date / Accepted: date}

\maketitle

\begin{abstract}
We present a novel method to reconstruct complex network from partial information. We assume to know the links only for a subset of the nodes and to know some non-topological quantity (fitness) characterising every node. The missing links are generated on the basis of the latter quantity according to a fitness model calibrated on the subset of nodes for which links are known. We measure the 
quality of the reconstruction of several topological properties, such as the network density and the degree distribution as a function of the size of the initial subset of nodes. 
Moreover, we also study  the resilience of the network to distress propagation. We first test the method on ensembles of synthetic networks generated with the Exponential Random Graph model which allows to apply common tools from statistical mechanics. We then test it on the empirical case of the World Trade Web. In both cases, we find that a subset of 10 \% of nodes is enough to reconstruct the main features of the network along with its resilience with an error of 5\%. 

\keywords{Complex Networks \and Financial Systems}
\PACS{89.75.Da \and 02.50.Le \and 89.65.Ef}
\end{abstract}

\section{Introduction}
\label{intro}

The reconstruction of the system from partial information is one of the outstanding and unresolved problems in the field of statistical physics of networks \cite{clauset2008missing-links,mastromatteo2011reconstruction}. Indeed, there are several real-world economic and financial contexts where knowledge of the entire network structure would be crucial to assess the resilience of the system to both exogenous and endogenous shocks while, at the same time, only limited information on that structure is available. An example is the case of financial networks where nodes represent financial institutions and links represent financial ties of various types such as loans or derivative contracts. These ties result in many cases in dependencies among institutions and constitute the ground for the propagation of financial distress across the network. The resilience of the whole system to the default or the distress of one or more institutions depends on the topological structure of the network \cite{battiston2012liaisons,battiston2012debtrank}. Unfortunately, due to confidentiality issues banks do not disclose their mutual exposures.

Typically the analysis of systemic risk is done by reconstructing the network using the so-called Maximum Entropy (ME) algorithm. This method assumes that the network is fully connected (for this reason this class of approaches is called "dense reconstruction methods").  The weights of the links are then obtained via a "maximum homogeneity" principle. This means that each node is assumed to bear a similar level of dependence from all other nodes. The second step consists in finding a matrix that, while satisfying certain constraints (imposed in this case by the budget of the individual banks), minimizes the distance from the uniform matrix in which each entry has the same value. Such a matrix is found using the Kullback-Leibler divergence as the objective function to minimize \cite{degryse2007interbank,lelyveld2006interbank}. The hypothesis that the network is fully connected is a strong limitation of the ME algorithm, since empirical networks are characterized by heterogeneous degree. Moreover, \cite{mastromatteo2011reconstruction} has shown how the "dense reconstruction" leads to an underestimation of the systemic risk. They also provide a new algorithm that allows to minimize Kullback-Leibler divergence obtaining a matrix with an arbitrary level of heterogeneity under a maximum value depending on constraints. Their algorithm provides a "sparse reconstruction" that is more reliable than the dense one.  Nevertheless it leaves open the question of what value of heterogeneity would be appropriate to choose, since the density of connections must be specified ex-ante and it is not recovered by the algorithm. 

In this paper we introduce a new general method, the Bootstrapping Method (BM), to reconstruct the topology of the whole network starting from the knowledge 
of a subset of nodes. This method overcomes some of the limitations described above. To validate the method we use both synthetic networks as well as an example of real economic systems. In both cases we compare the reconstruction with the known total structure. This is the so-called  World Trade Web (WTW), i.e. the network of countries where links represent the financial flows corresponding to the trade volumes among them. In our method, the allocation of the links among nodes is carried out using the fitness model\cite{caldarelli2002fitness,caldarelli2007fitness}. Differently from other network generation models, the fitness model generates a network structure starting from a non-topological variables (fitness) associated to the nodes. This approach has been used in the past to reproduce the topological properties of several empirical economical networks, including the network of equity investments in the stock market \cite{garlaschelli2005investment}, the interbank market \cite{demasi2006fitness-interbank}, the currency market \cite{Galluccio1997currency}, and the  WTW \cite{garlaschelli2004fitness-wtw}. We investigate how well it is possible to recover both the topological properties of the network and its resilience to distress propagation, as we vary the size of the subset of nodes for which information is available. Among the topological properties, we focus on those that play an important role in contagion processes and in the propagation of distress, i.e., the network density \cite{battiston2012liaisons}, the degree distribution \cite{pastor2001epidemic},  the k-core structure\cite{kitsak2010identification}. 

We find that having information on a relatively small fraction of nodes is sufficient to recover with good approximation the above properties. For instance, with  only about 7\% of the nodes (10 out of 185) we have a relative error of about: 7\% on the density, 10\% on the average degree of the main core, 7\% on the size of the main core. For the resilience, we focus on a recently introduced notion, debtrank \cite{battiston2012debtrank}, which measures the systemic impact of the initial distress of a subset of nodes, whenever the links in the network represent the dependencies among nodes. Similarly to the above results, we find that with about 7\% of the nodes the resilience is recovered with a relative error within 10\%. 

At a first thought, it can be surprising that a small fraction of nodes enables to reconstruct so well global emerging properties of the network. However, one should bear in mind that in the method, while the links are known only for a subset of nodes, the fitness is always known for all the nodes. 
Thus, the method would probably require much higher fraction of nodes in order to reconstruct networks with special topologies such as strong community structure or networks where the fitness is not a strong factor in driving the connectivity. The investigation of these effects is left for future research. Overall, our method can be applied in   principle to any network representing a set of dependencies among components in a complex system and it is thus of general interest in the field of complex networks and statistical physics.

\section{Exponential random graph and fitness model}\label{sec:expon-rand-graph}

In this paper we propose a Bootstrap Method (BM) to build the network using a fitness model. The method is described in Section \ref{sec:bootstrapping-method}. Here we briefly describe the ERGMs and the associated fitness model. In order to generate ensembles of complex networks both dynamic and static approaches can be utilized.  In the dynamic case,  nodes and/or links are added step by step using for instance a "preferential attachment" algorithm. In the static case,  instead, the number of nodes is fixed the links are assigned at once according to some  statistical or deterministic criterion.
Exponential random graph models (ERGM) are one of the most studied class of network models\cite{park2004statistical,dorogovtsev2010lectures}. They can be described using the powerful mathematical formalism of the equilibrium statistical mechanics \cite{park2004statistical}.  

As a specific example, we will consider the so-called \textit{fitness} or \textit{hidden variables} models, where the network topology is determined by an intrinsic property (called \textit{fitness}) associated with each node of the network\cite{caldarelli2002fitness}.  Through this scheme we can define a framework to investigate those networks where the topology is driven, at least in part, by  non-topological properties of the nodes. With the fitness model it is possible to study several economical networks, ranging from the  WTW (where the fitness of the model are the GDP of the various countries) \cite{garlaschelli2004fitness-wtw},  to the financial networks (where fitness are, for instance, the market capitalization of each  institution) \cite{garlaschelli2005investment,demasi2006fitness-interbank}. 
 
Given a set of network properties, $\{C_a\}$ the Exponential Random Graph Model (ERGM) is defined as the ensemble $\Omega$ of maximally random networks with $\{C_a\}$ constrained to some statistical properties. More specifically, let us suppose that the ensemble averages of $\{ C_a \} $ are fixed:
\begin{equation}
 \langle C_a \rangle_{\Omega} \equiv \sum_G P\,(G) C_a(G) = C_a^* ~~~~~~ \forall a \label{eq:constraints}
\end{equation}
It has been shown that $\Omega$ can be defined through a set of control parameters $\{\theta_a\}$, the values of which depend on a set of constraining values $\{C^*_a\}$ \cite{park2004statistical,dorogovtsev2010lectures}.  Furthermore the probability $P\,(G)$ of a network $G$ to occur in $\Omega$ is given by $P\,(G)\,=\,e^{-H(G)}/Z$, where we introduced the graph Hamiltonian $H\,(G) \equiv \sum_a \theta_aC_a(G)$ , and the partition function $Z \equiv \sum_G \exp(-H\,(G))$.  $\{\theta_a\}$ is the set of Lagrange multipliers associated to the constraints $\{C^*_a\}$.
The fitness model can be seen as a specific case where the set of properties $\{C_a\}$ is the degree sequence $\{k_i\}_{i=1,...N}$ of the nodes of the network. 
In this case $H=\sum_i \theta_i k_i$, the partition function is exactly computable and each node can be identified by its control parameter (or Lagrange multiplier) $\theta_i$. Fixing the values of $\{\theta_i\}$ is equivalent to fix the mean values of
$\{k_i\}$. In order to  further clarify the role of $\{\theta_i\}$ in controlling the topology, let us define $x_i \equiv e^{-\theta_i}$. It is possible to show that the ensemble is such that for each network in $\Omega$ two nodes $i$ and $j$ are connected with a probability given by: 
\begin{equation}
 p_{ij}=\frac{x_ix_j}{1+x_ix_j} \label{eq:prob linking}.
\end{equation}
Therefore $x_i$ can be considered as the fitness of the node $i$ and it is related to the ability of $i$ to create links with other nodes.

The average in $\Omega$ of several topological properties of the network can be expressed in terms of appropriate compositions of the linking probabilities $p_{ij}$ for every $i$ and $j$. For instance, we can write the degree $k_i$ as 
\begin{equation}
 \langle k_i \rangle = \sum_{j\,(\neq i\,)=1}^{N}p_{ij} ; \label{eq:k_i vs p_ij}
\end{equation}
the average nearest neighbor degree $K^{nn}_i$ as 
\begin{equation}
 \langle k_{i}^{nn}\rangle=\frac{\sum_{j\neq i}\sum_{k\neq j}p_{ij}p_{jk}}{\langle k_{i}\rangle} ;
\end{equation}
and the clustering coefficient $C_i$ as 
\begin{equation}
 \langle C_{i}\rangle=\frac{\sum_{j\neq i}\sum_{k\neq j,i}p_{ij}p_{jk}p_{ki}}{\langle k_{i}\rangle\left[\langle k_{i}\rangle-1\right]}.
\end{equation}
In the limit of small values of fitnesses (and therefore small connectivity), $x_i$ 
is proportional to the \textit{desired degree} of the node $i$.  Indeed, in this limit we can assume $\langle k_i \rangle \simeq \sum_j x_ix_j \propto x_i$.


\section{Bootstrapping Method}\label{sec:bootstrapping-method}
The estimation of the linking probability $p_{ij}$ between node $i$ and node $j$,  $p_{ij}$ is the initial step in order to develop a network Bootstrapping method.  Let us suppose to have incomplete information about the topology of a real network (say $G_0$). In particular, we assume to know the links of only a subset $I$ of the nodes. Moreover, we assume to know, for all the nodes, a non-topological property, denoted as $y_i$, that is correlated to  some statistical properties of the degree $k_i$ of the nodes by a  known relation as below clarified. For instance, in the World Trade Web $y_i$ could be the country GDP, while in financial networks it can be the operating revenue of the firm $i$. We would like to estimate the value $x\,(G_0)$ of a topological property $X\,(G_0)$ of the network $G_0$. We make two hypotheses:

\begin{enumerate}
 \item the network $G_0$ has been drawn from an ensemble of ERGM, that we call $\Omega$. From the statistical mechanics of networks we know that the value $x\,(G_0)$ of the property $X$ in $G_0$, varies within the range $\langle x \rangle_{\Omega} \pm \sigma_{x}^{\Omega}  $  where $\sigma_{x}^{\Omega}$ is the standard deviation, and  $\langle x \rangle_{\Omega}$ the average of the property $X$ estimated on the whole ensemble $\Omega$.

 \item each known value of the non-topological property $y_i$ is assumed to be proportional to the fitness, denoted as $x_i$ (because a generic property of the network can be used as a fitness variable) of the node $i$ in the ensemble $\Omega$, through a universal unknown parameter $z$: $\sqrt{z}y_i=x_i$ . Therefore Eq. (\ref{eq:prob linking}) becomes: 
\begin{equation}
 p_{ij}=\frac{zy_iy_j}{1+zy_iy_j} \label{eq:prob linking z}
\end{equation}
\end{enumerate}

With these hypotheses, we map the problem of evaluating $x\,(G_0)$ into the one of choosing an ERGM ensemble $\Omega$ compatible with the constraints given by the fitness. Once $\Omega$ is determined (it is univocally defined by the set of $\{x_i\}$), we can use the average $\langle X \rangle_{\Omega}$ as a good estimation for $x\,(G_0)$. Within this framework, the question is which ensemble $\Omega$, belonging to the class ERGM, is the most probable to extract the real network $G_0$, knowing only the partial information $\{y_i\}$. Since we know $\{y_i\}$ , i.e. the rescaled fitness values (a non topological property of the network), the problem becomes to find the most likely value of $z$.  For this reason we use the notation $\Omega(z)$  for the desired ensemble. 

Notice that if we knew not only $\{y_i\}$ but the entire topology of the network, $z$ could be found by means of a maximum likelihood argument (Ref. \cite{garlaschelli2004fitness-wtw}) comparing the average number of links in the ensemble networks with the total number of links $L_0$ in $G_0$: 
  \begin{equation} 
 \langle L \rangle = \frac{1}{2} \sum_{i=1}^{N} \langle k_i \rangle = \frac{1}{2} \sum_{i=1}^{N} \sum_{j \neq i} p_{ij} = L_0 \label{eq:L} 
 \end{equation} 
where $p_{ij}$ contains the unknown parameter $z$ through Eq. (\ref{eq:prob linking z}). Therefore we can evaluate $z$ as $L_0$ is known and by definition of $\sqrt{z}y_i = x_i $ go back to $x_i$ that is our desired output. Let us call $z_0$ the estimation calculated in this way, and $\Omega(z_0)$ the respective ERGM ensemble. But note that we assume to know only the degrees of the nodes in a subset $I$ and not the entire topology.  Let be $n$ the number of nodes of $I$. In this case the relation we have to apply in order to use the maximum likelihood principle in the estimation of $z$ is:
\begin{equation}
 \sum_{i \in I} \langle k_i \rangle = \sum_{i \in I} \sum_{j \neq i} p_{ij} = \sum_{i \in I} k_i
  \label{eq:estimate z}
\end{equation}
where the degrees $k_i$ are calculated in the original network $G_0$. For a subset $I$ of the entire set of nodes of the network the estimation is less precise than the one in Eq. (\ref{eq:L}). However even with just the knowledge of the degree of a single node, the  Eq. (\ref{eq:estimate z}) estimates $z$, and finally $X\,(G_0)$ .

The network bootstrap of a network $G_0$ is defined by the above equations using the following procedure. Let us assume to know the non topological property $y_i$  of all $N$ nodes of the system and the links of a subset $I$ of $n < N$ nodes. 
\begin{itemize}
\item Given the topological information of the links in the subset $I$, we compute the sum of all degrees of these $n$ nodes in $G_0$: $\sum_{i \in I} k_i$ .
\item This sum is substituted into the Eq. \ref{eq:estimate z} to obtain the relative value of $z$, denoted as $z'$, that is an approximation of the $z_0$.
\item With the value of $z'$ and the knowledge of every $y_i$ we assign all the links in the network according to the linking probability of Eq. \ref{eq:prob linking z}.
\end{itemize}

We want to estimate the accuracy of the network bootstrap method for both topological and non-topological properties. To this end, we first apply the method to a synthetic network generated using the fitness model (see Section \ref{sec:test-bm:-synthetic}). We then apply the method to an empirical case, i.e. the WTW (see Section \ref{sec:test-bm:-world}). In the second case, we test also how well we can reconstruct a global and non-topological property such as the resilience of the network to distress propagation  (see Section \ref{sec:test-bm:-debtrank}).

\section{Test of BM: synthetic networks}\label{sec:test-bm:-synthetic}


Let be $\{I_{\alpha}\}$ , with $\alpha = 1,...,M$, an ensemble of subsets of the network $G_0$ each of them  containing $n$ nodes. In order to test how much our BM estimate of the property $X$ is precise, we will proceed in the following way:
\begin{itemize}
 \item evaluate $z$ for each  subset $I_{\alpha}$ from Eq. (\ref{eq:estimate z}), let us call it $z_{\alpha}$
 \item  use the value $z_\alpha$ to estimate, through the relation $\sqrt{z}y_i = x_i$, the average property $\langle X \rangle_{\alpha} \equiv \langle X \rangle_{\Omega(z_{\alpha}) \equiv I}$  
 \item repeat the calculation for all other sets $I_\alpha'$, accumulate the values of $\langle X \rangle_{\alpha}$ and compute the average $ \overline{ \langle X \rangle}$ of this quantity and its associated root mean square deviation with respect to the real value of $X(G_0)$ across all the realizations of $I_\alpha$, for fixed $n$.
\end{itemize}
The property $X$ is then estimated by averaging the $\langle X \rangle $ computed for each subset $I$. Notice that each value $\langle X \rangle $ is by itself an estimation of the true, unknown, property $X$.

In order to study the accuracy of the reconstruction, we study how the root mean square error varies as a function of the size $n$ of the subset of nodes for which information is available. Using  the fitness model and all the available information, we generate an ensemble of networks $G$ each one of size $N$ and we compute several  properties like the network density, the size of the main core and the average degree of the main core. These values will be our benchmarks to test how good is the network reconstruction with the BM.
 
We test the BM  by using the following three topological quantities because they have been found to play a role in the distress propagation and contagion processes and therefore are relevant to the resilience of the network to systemic risk (see Section \ref{intro}):
\begin{enumerate}
 \item \textbf{density} $D$ ;
 \item \textbf{degree of the main core}, $k^{main}$. In a network, the $k$-core is defined as the ``largest subgraph whose nodes have at least $k$ connections (within this subgraph, of course)'' \cite{dorogovtsev2010lectures}. The main core is $k$-core with the highest possible degree, $k^{main}$ ;
 \item \textbf{size of the main core}, $S^{main}$ , i.e. its number of nodes.
\end{enumerate}

Each of these measures will play the role of the property $X$ in the previous notation. In order to use a real-world fitness, we take as reference the WTW (in year 2000) which contains 185 nodes. We thus generate networks of size N=185 and we use as fitness $y_i$ the GDP from the WTW. For each of these properties we will carried out the procedure described here below.

\begin{figure}[ht!]
\begin{center}
\begin{tabular}{cc}
 \includegraphics[scale=0.3]{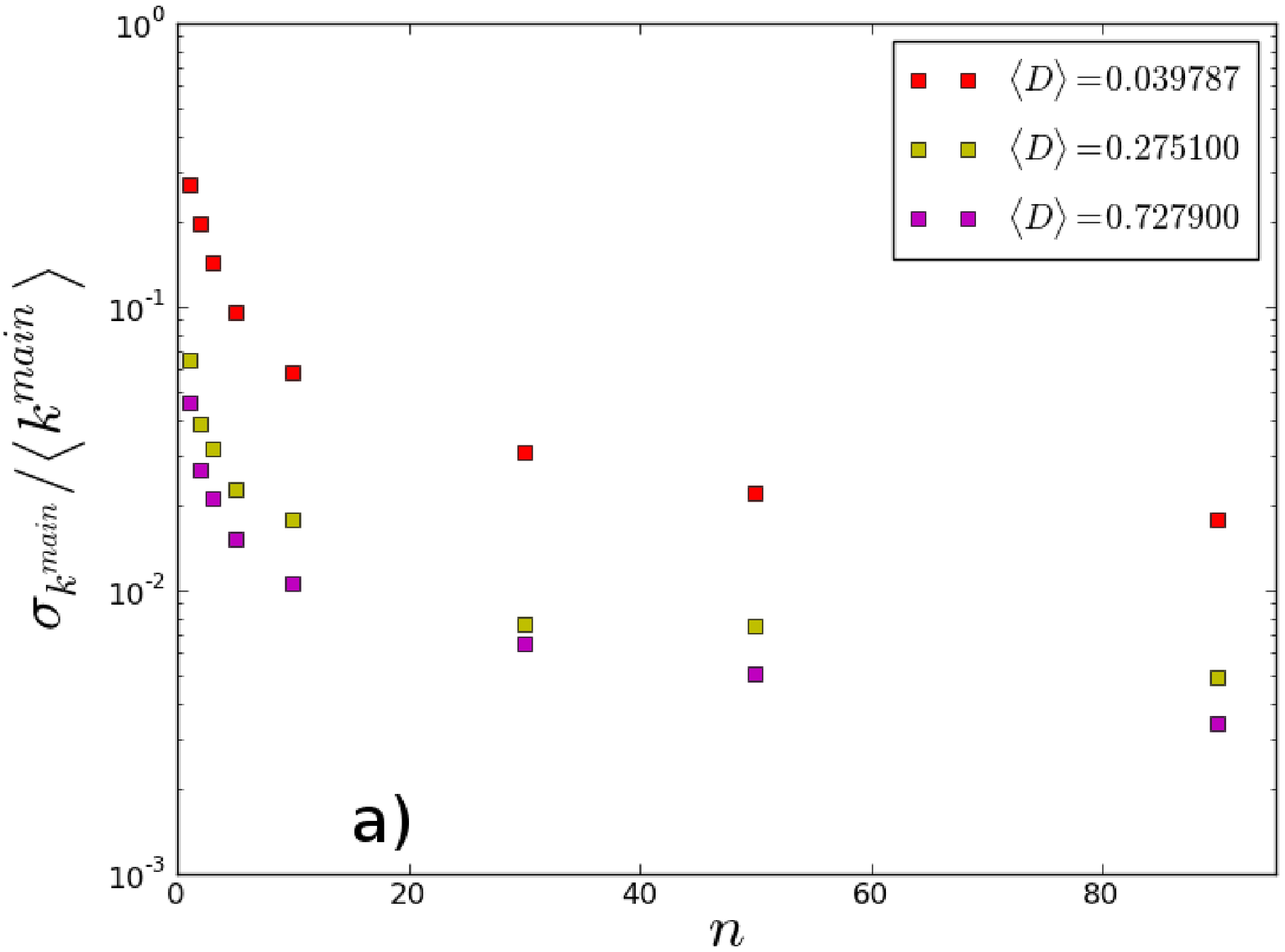} &
 \includegraphics[scale=0.3]{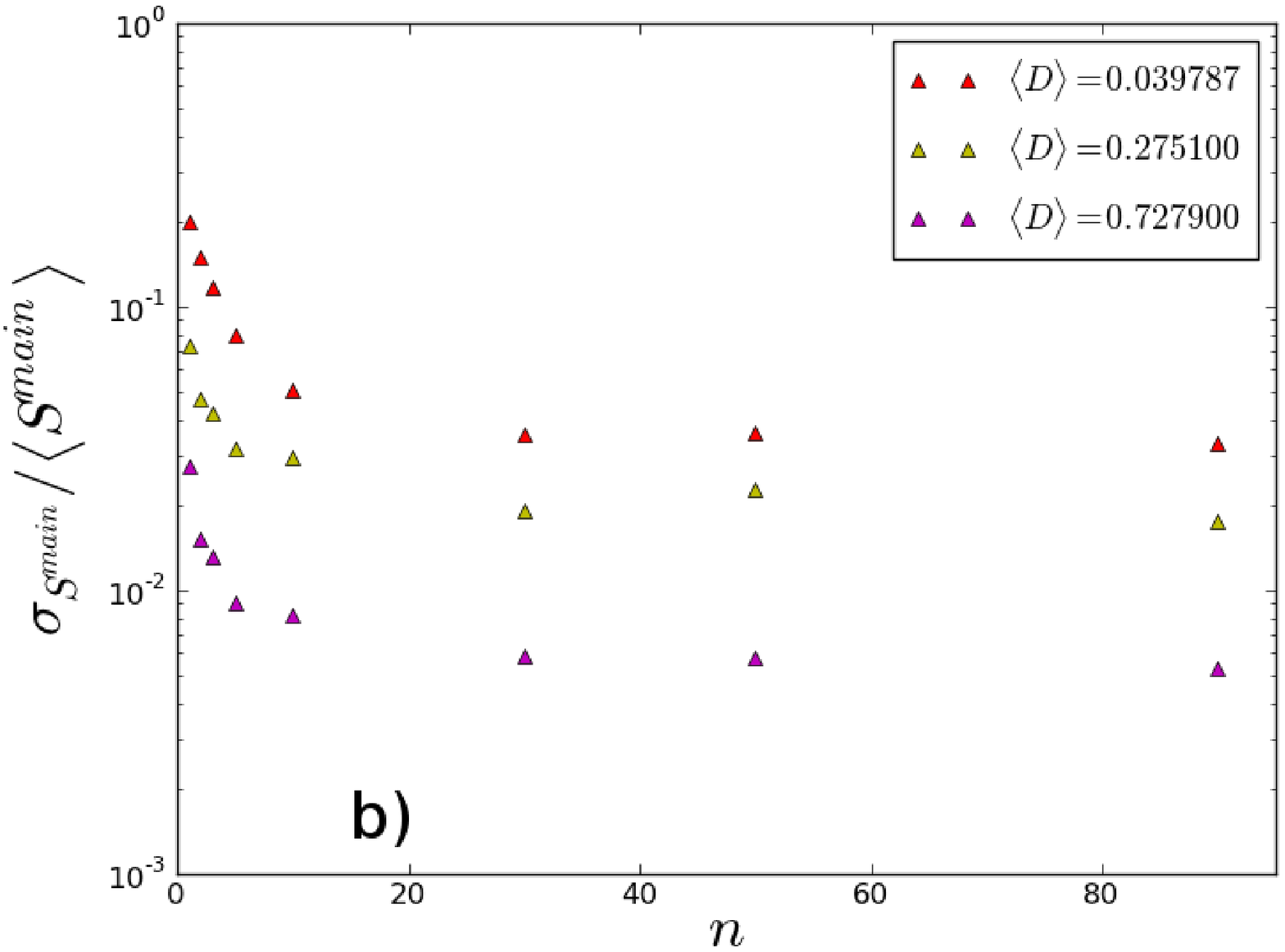} \\
 \multicolumn{2}{c}{ \includegraphics[scale=0.3]{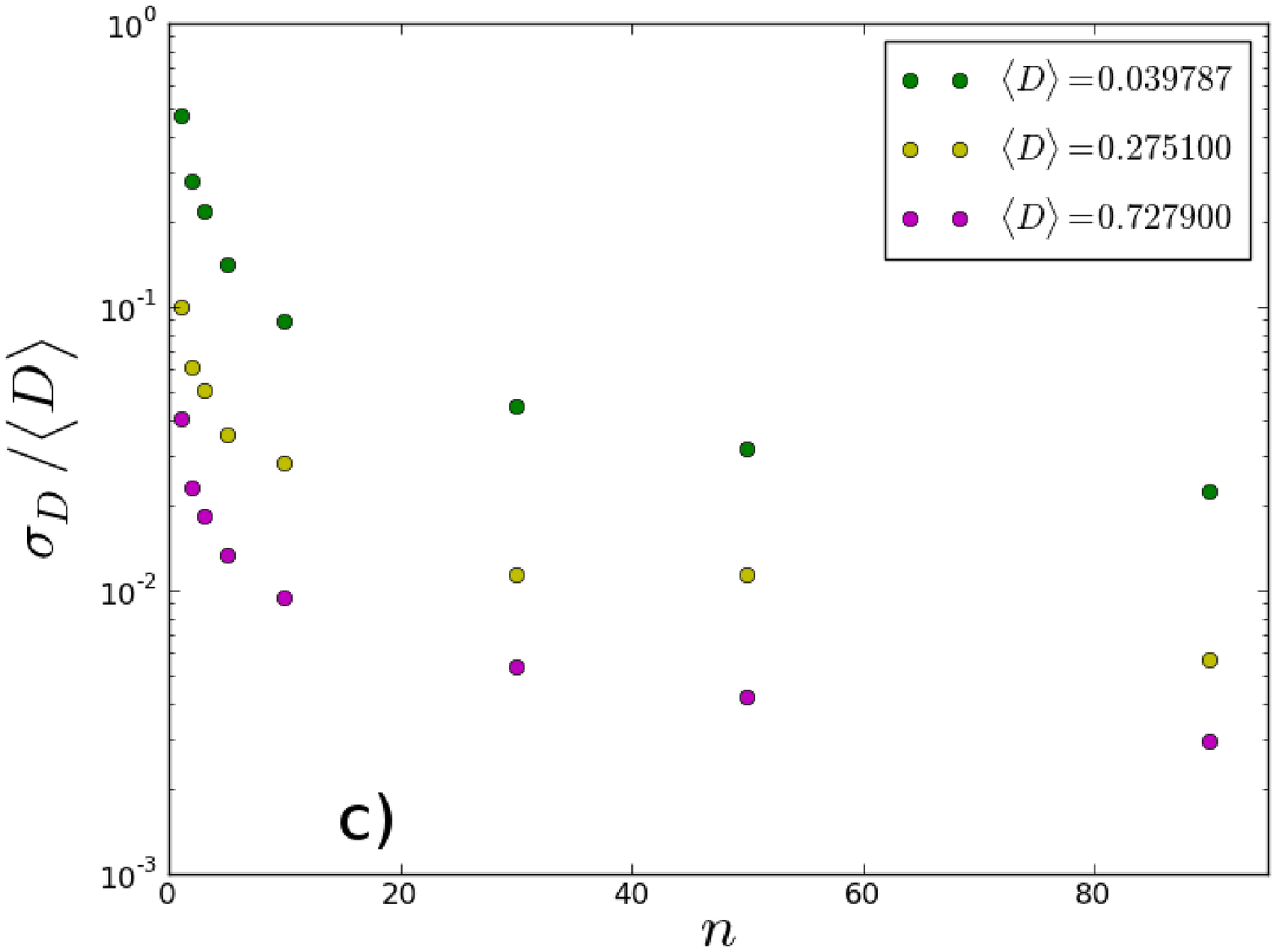}}
 \end{tabular}
\end{center}
\caption{\label{fig:test_on_k_S_D_variousZ} The pictures from top left represent respectively:  a) using three values of $z_0$  related to three different link average densities $\langle D \rangle $ (estimated numerically) compute the relative error $\sigma_k^{main} / k_{e}^{main}$ for the three cases b) same as (a) but for the relative error of the \textit{S}-main core size, c) same as in (a) but for the density of the links $D$. In all the 3 plots it is evident how the  quality of the reconstruction increases with the number of nodes used to generate the network ensemble.}
\end{figure}

\begin{enumerate}
\item choose a value for the  variable $z_0$ (compatible with the fitness model for WTW, where the fitness is the GDP of a country); we start with $z_0 = 10^4$
\item using as fitness the GDP of a country create 50 networks. Let be $\Omega_N$ this ensemble. Compute on the this set the average link density $D_e$
\item use a 51th network from the $\Omega_N$ ensemble as reference network, call it $G_0$, this will be the network to reconstruct
\item starting from network with a single node $n = 1$ with known degree $k$ and GDP $y_i$ use this information to compute an estimation of $z$, say $z'$
\item from the new value of $z'$ create a new ensemble of 50 networks (say $I_\alpha$ because  is referring to one particular set of random chosen nodes)
\item choose another set of $n$ nodes, generate 50 networks from this set, repeat this operation 100 times each time with a different set $I_\alpha$ of $n$ nodes 
\item in each of the 100 ensembles of 50 networks, $I_\alpha$ estimate the average density $\langle D\alpha \rangle$
\item compute the root mean square error: 
$ \sigma_d =  1/100 \sum_{\alpha} \sqrt{ \left( \langle D_{\alpha} \rangle - D_e \right)^{2}}$ 
the difference is between the reconstructed networks $\langle D_{\alpha} \rangle$ and the original average link density $D_e$
\item compute and plot $\sigma_d / D_e$
\item repeat the points from 4 to 9 using a different values of $n$  
\end{enumerate}
The entire procedure is repeated for the quantities $S^{main}$ and $k^{main}$, and the results are shown in Fig.\ref{fig:test_on_k_S_D_variousZ} for 3 different values of $z_0$, corresponding to different values of density. 
We observe that in all cases there is a rapid decrease of the relative error as the number of nodes $n$, used to reconstruct the topology, increases. This is a good indication of the goodness of the method. Even with a single node, plus the information on the fitness $y_i$ (GDP of the countries), we are able to estimate the topological properties of the network with a relative error of about 13\% for the main core average degree $k^{main}$, about 18\% for the network density $D$, and about 10\% for the size of main core $S^{main}$.

As expected, if we have a denser network (Fig.\ref{fig:test_on_k_S_D_variousZ}d-f ) the relative error is smaller because the network has more links from with the BM can reconstruct the topology. The same trend in the decrease of the relative error is verified for all the examined topological quantities.  

\newpage

\section{Test of BM: World Trade Web}\label{sec:test-bm:-world}
We now test the empirical network of the WTW for the same topological properties of the previous case. The main difference is that now instead of using a reference network generated with the fitness model and an average measure of this network class (generated in the ensemble $\Omega_N$ ), the reference is now the empirical WTW network.
We perform the test with the following similar procedure:
\begin{enumerate}
\item compute the variable $z_0$ from the WTW using the GDP of the countries and all the links of the original network
\item from the WTW network compute the density $D_{WTW}$ of the links
\item starting from a network with a single node $n = 1$ with known degree $k$ and GDP $y_i$ use this information to compute an estimation of $z$, say $z'$
\item from the new value of $z'$ create a new ensemble of 50 networks (say $I_\alpha$ because  is referring to one particular set of random chosen nodes)
\item choose another set of $n$ nodes, generate 50 networks from this set, repeat this operation 100 times each time with a different set $I_\alpha$ of $n$ nodes 
\item in each of the 100 ensembles of 50 networks, $I_\alpha$ estimate the average density $\langle D\alpha \rangle$
\item compute the root mean square error: $ \sigma_d =  1/100 \sum_\alpha \sqrt{\left(\langle D_{\alpha} \rangle - D_{WTW}\right)^{2}}$ the difference is between the reconstructed networks $\langle D_{\alpha} \rangle$ and the original WTW link density $D_{WTW}$
\item compute and plot $\sigma_d / D_{WTW}$
\item repeat the points from 4 to 9 using a different value of $n$  
\end{enumerate}

the same test is carried out for the other quantities $k^{main} and S^{main}$. We expect to see more error than the previous case. We plot the quantity 

\begin{figure}[ht!]
\begin{center}
\begin{tabular}{cc}
 \includegraphics[scale=0.3]{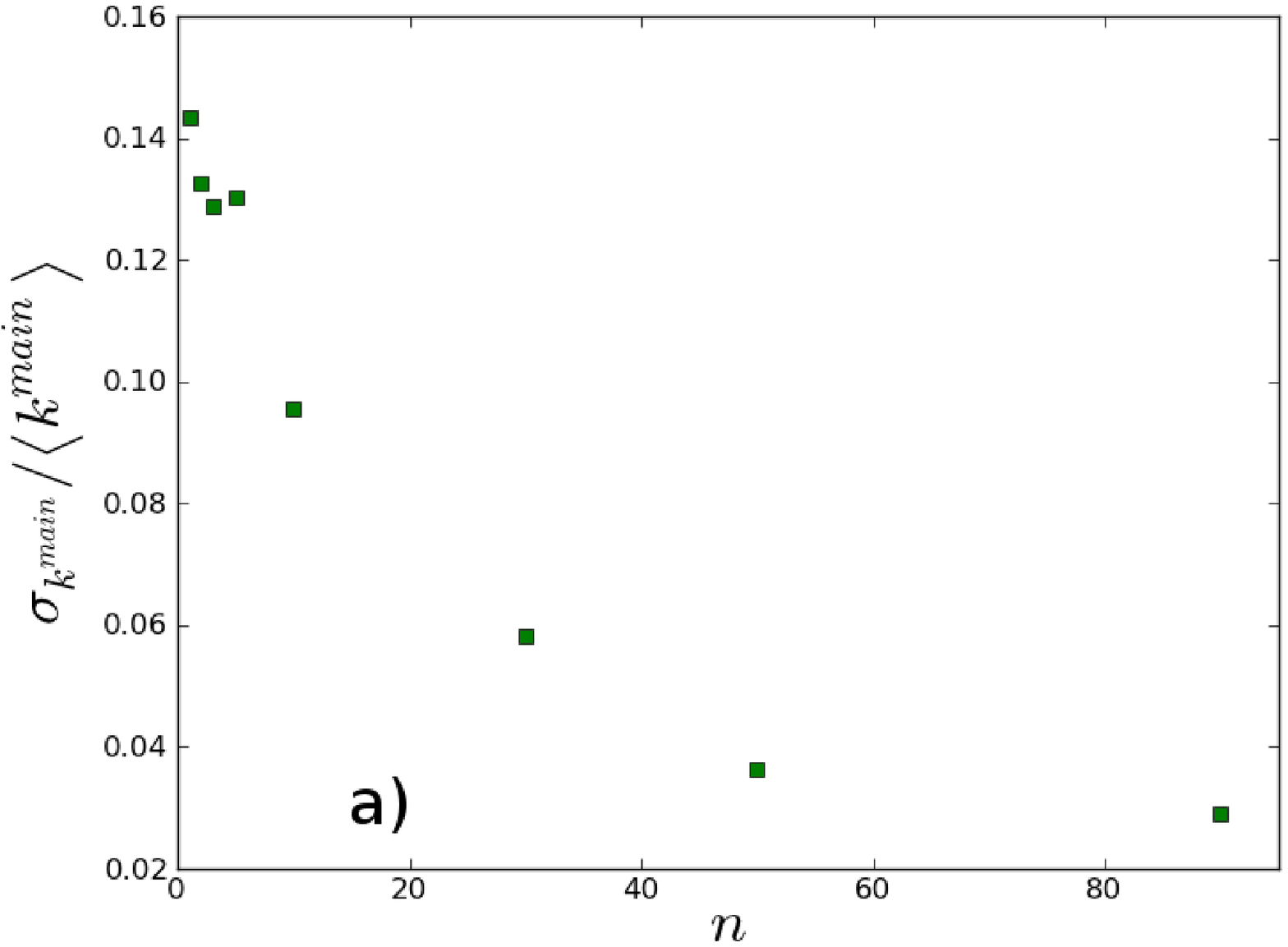} &
 \includegraphics[scale=0.3]{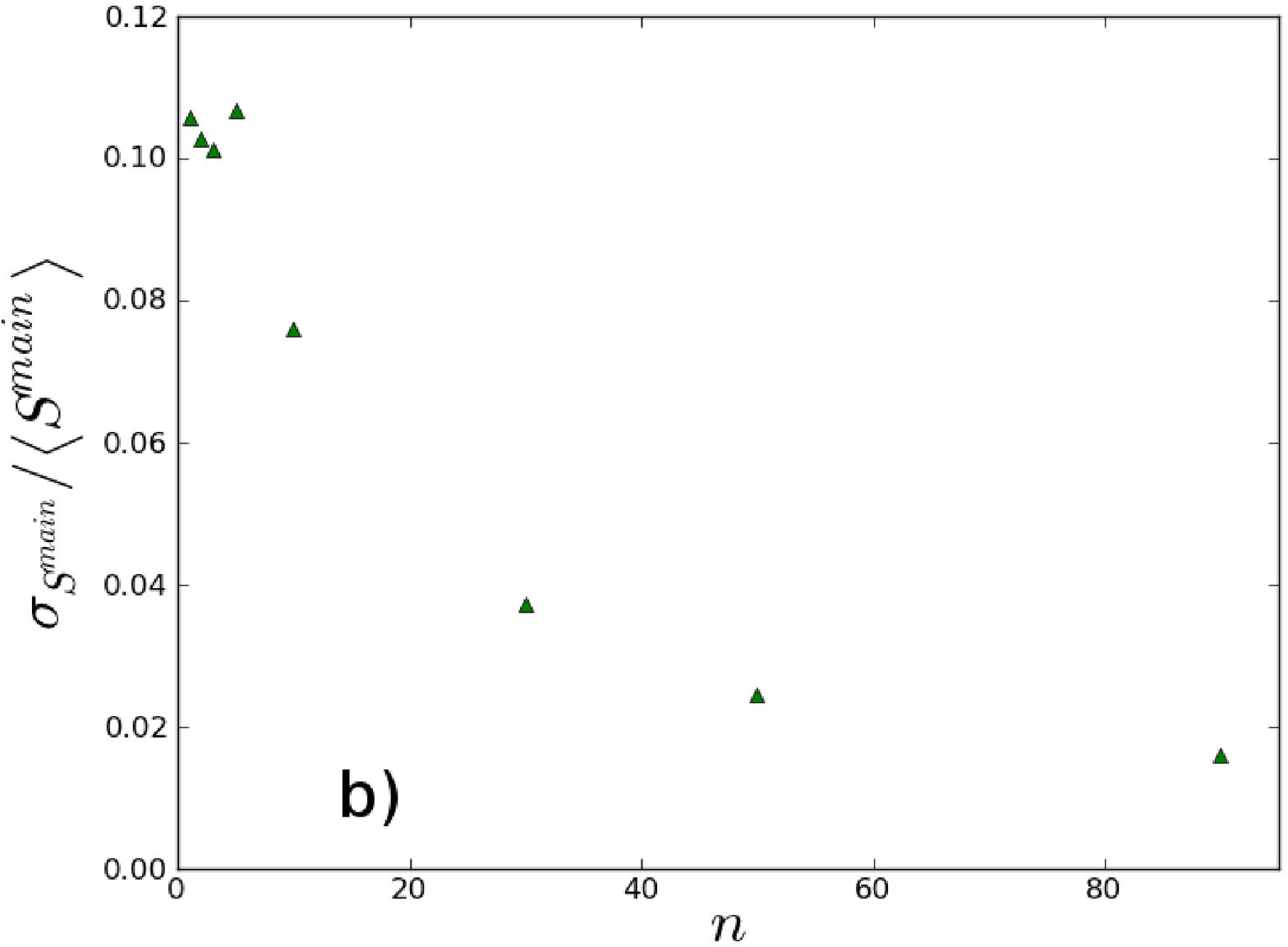} \\
  \multicolumn{2}{c}{\includegraphics[scale=0.3]{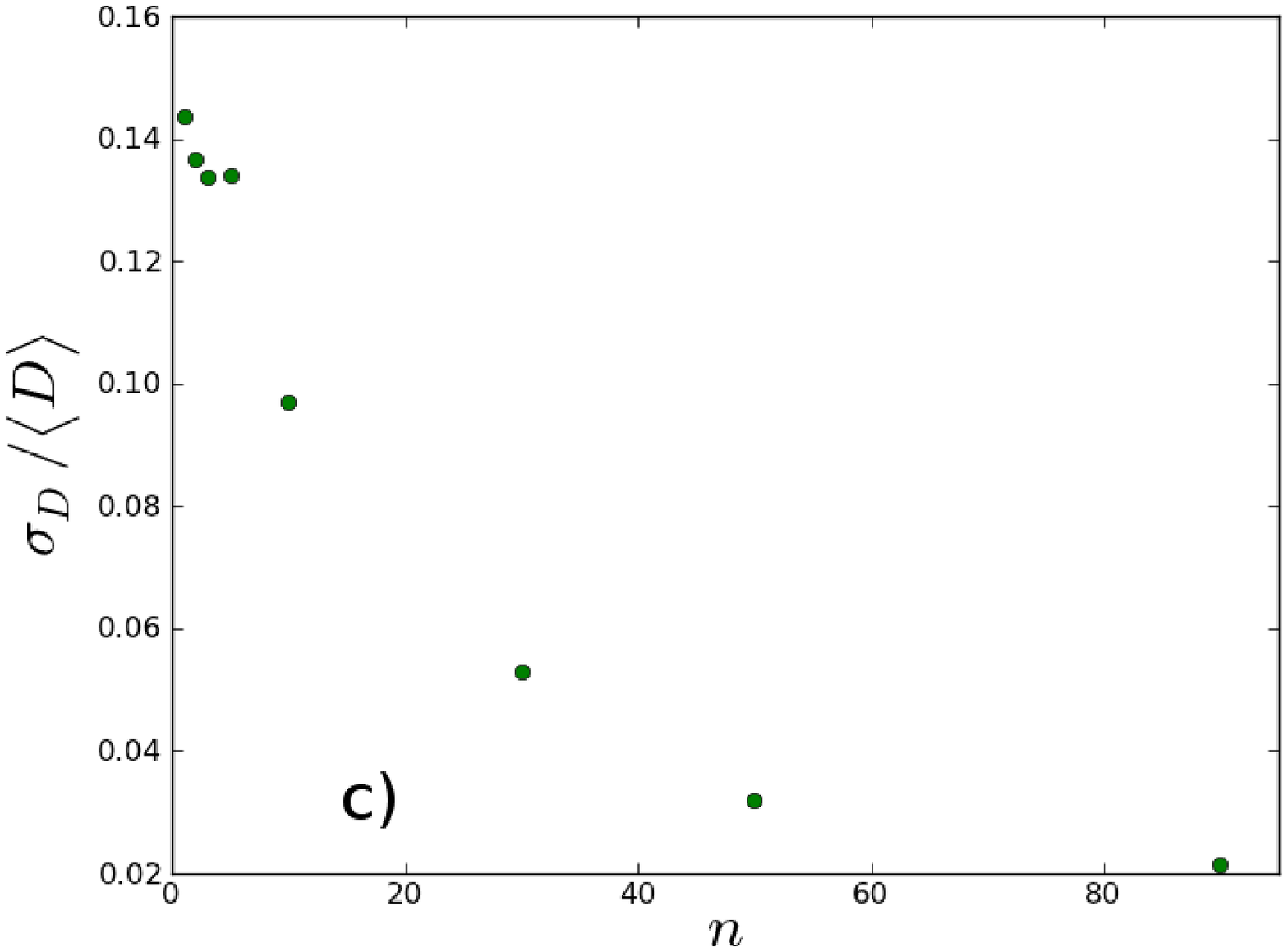}} 
 \end{tabular}
\end{center}
\caption{\label{fig:test_on_k_S_D_WTW} The pictures from top left represent respectively: a) the relative error in the estimation of average degree of the main core $\sigma_k^{main} / k_{WTW}^{main}$ computed with real WTW network following the procedure described in the previous paragraph b) same as in (a) but for the relative error in the size of main core c) same as in (a) but for the density of the links $D$. In all the 3 plots it is evident how the goodness of the reconstruction of the WTW network increases with the number of nodes used to generate the network ensemble.}
\end{figure}

\section{Test of BM: DebtRank a measure of systemic risk}\label{sec:test-bm:-debtrank}

With the WTW network we can use a novel measure of systemic risk: the DebtRank \cite{battiston2012debtrank} that represents the expected distress of the nodes in case of financial events. In the WTW case a financial event can be, for instance, the default of a country and the subsequent impossibility to pay the traded goods: this \textit{shock} generate a distress propagation in the network causing losses to the other countries. The DR captures the impact $I_i$ of the shock to each node $i$. 

 We compute the DebtRank DR of a single node (due to a shock hitting a single node at a time), and the group debtrank GDR of the ensemble (due to an initial shock hitting all the nodes simultaneously) using an algorithm described  in the Methods section of  \cite{battiston2012debtrank} that consists in computing a feedback centrality from the matrix of the weights, given an initial shock $\psi$ (to one or more nodes) that is carried out by a variable $h_i$ \textit{impact} specific to the method. During the calculation of this feedback centrality we use several values of the impact rescaling factor $0 < \alpha < 1  $ to propagate the shocks in the network.

To compute the DR of each node we use the procedure:

\begin{enumerate}
\item choose an impact rescaling factor $\alpha$, the greater is this factor the greater will be reverberation effect on the network
\item assign an initial shock $0 < \psi < 1$ to a node $i$
\item run the DR algorithm (as described in the Methods section of \cite{battiston2012debtrank})
\item save the values of the impact at the end and the beginning of the simulation: the DR of the node $i$ will be the difference in the impact on all nodes after the propagation of the distress
\item repeat for a different node 
\end{enumerate}

\begin{enumerate}
\item impact rescaling factor $\alpha$
\item assign an initial shock $0 < \psi < 1$ to \textit{all} nodes
\item run the DR algorithm 
\item save the values of the impact at the end and the beginning of the simulation. The GDR will be the difference in the impact on all nodes after the propagation of the distress: $\label{eq: group-debt-rank} GDR = \sum_j h_{j}(T) v_j - \sum_j h_{j} (1) v_j,$ where $h_i$ is the impact on each node, $T$ indicates the end of the simulation (when the distress propagates to the entire network).
\end{enumerate}

Our goal is to test how well the GDR is computed with the network bootstrap. We make several tests for different values of initial impact $\psi$ and impact rescaling factor $\alpha$.  The DebtRank is strongly dependent by the weights (value of the elements of the adjacency matrix) that are instead unknown during the simulations. In fact the fitness model reconstructs the degree sequence, not the weight of the nodes we then use a value for each link with two rules:
\begin{itemize}
\item Compute the average weight averaging the elements of the $W_{ij}$ matrix associated to the $n < N$ nodes. Use this value  as \textit{homogeneous} weight for all nodes 
\item Assign to each node an weight similarly to a gravity model (see \cite{fagiolo2010gravity}) where the link $l_{ij}$ has a weight proportional to the product of the GDPs $GDP_i \cdot GDP_j$
\end{itemize}

\begin{figure}[ht!]
\begin{center}
\begin{tabular}{cc}
 \includegraphics[scale=0.3]{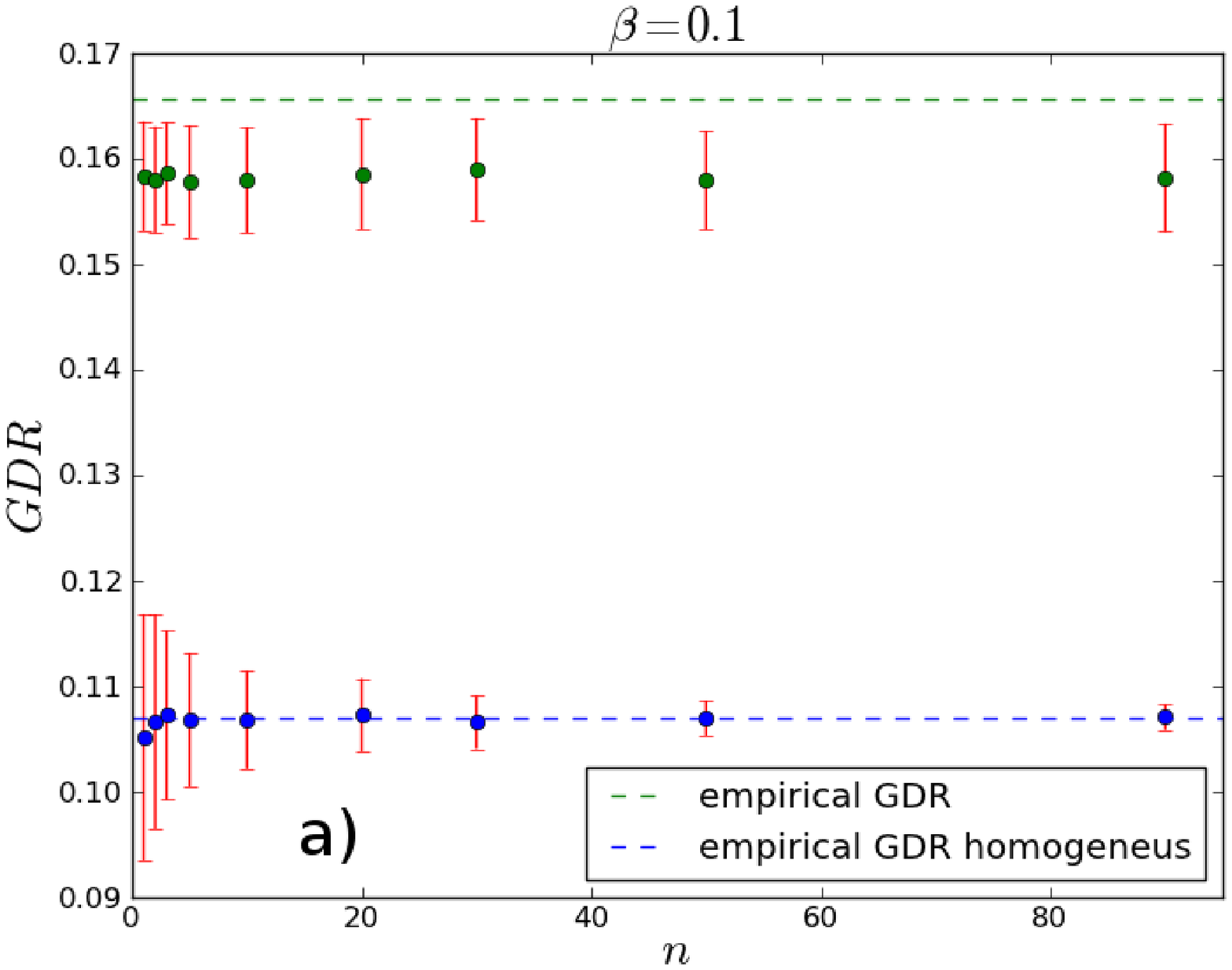} &
 \includegraphics[scale=0.3]{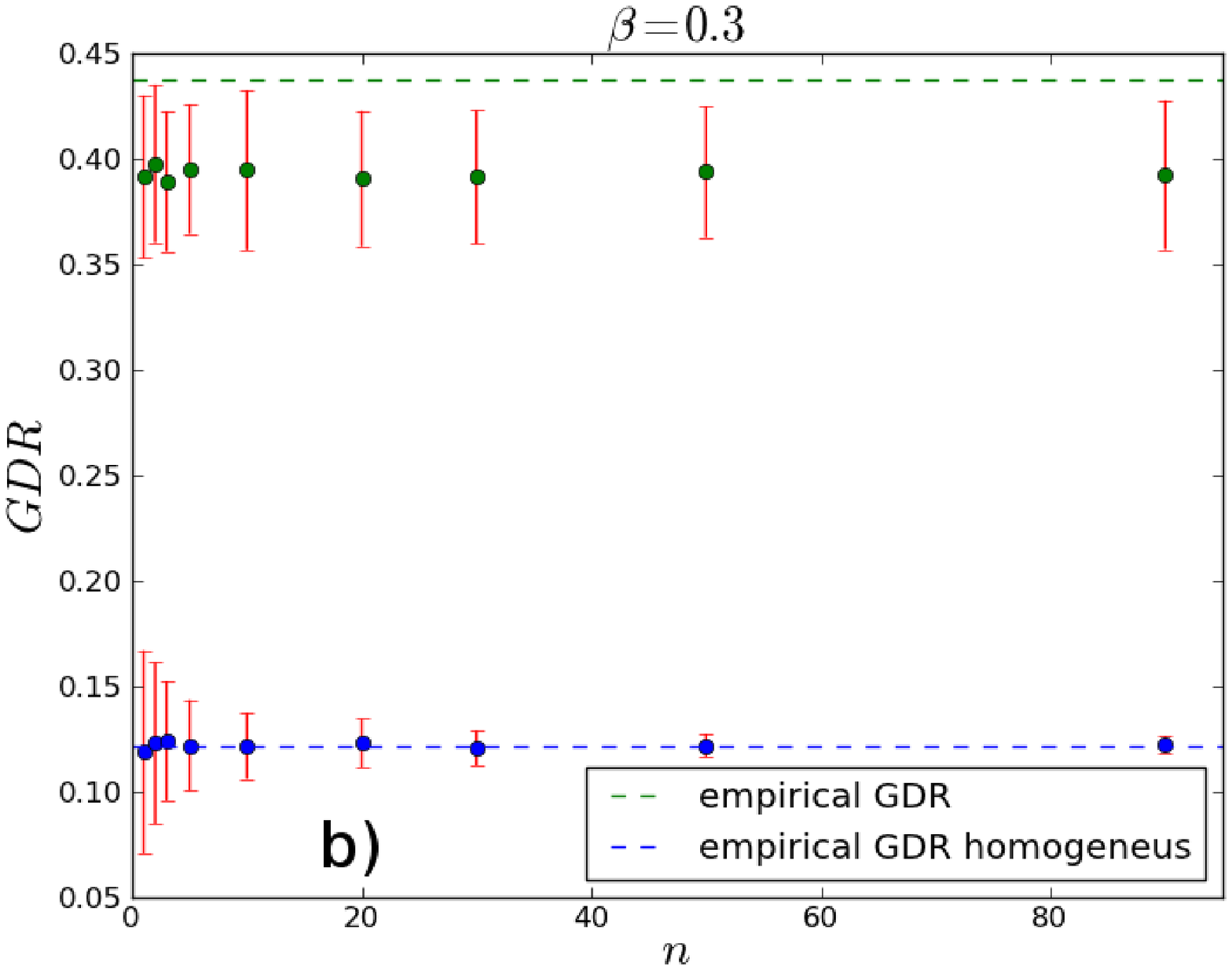} \\
 \includegraphics[scale=0.3]{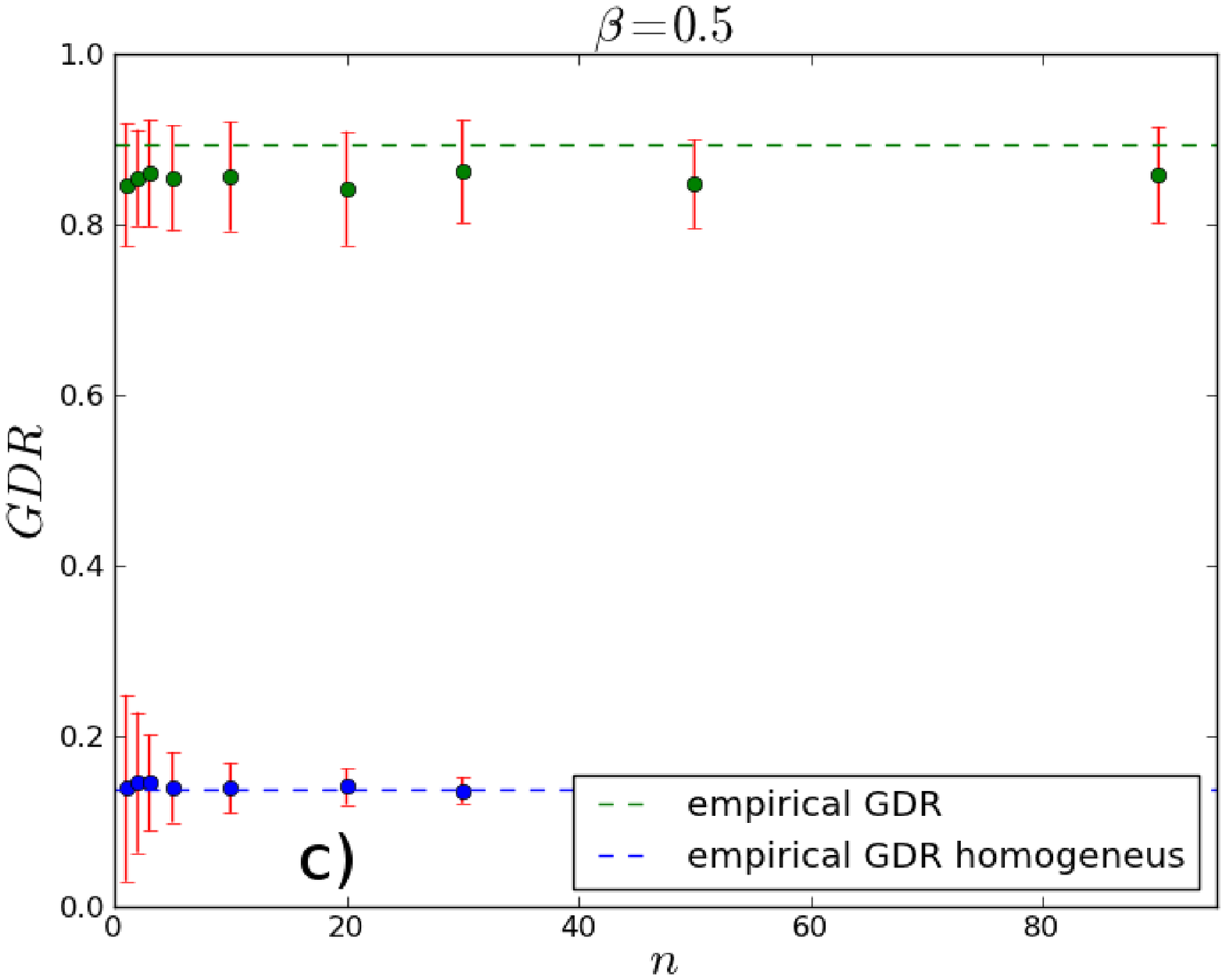} &
 \includegraphics[scale=0.3]{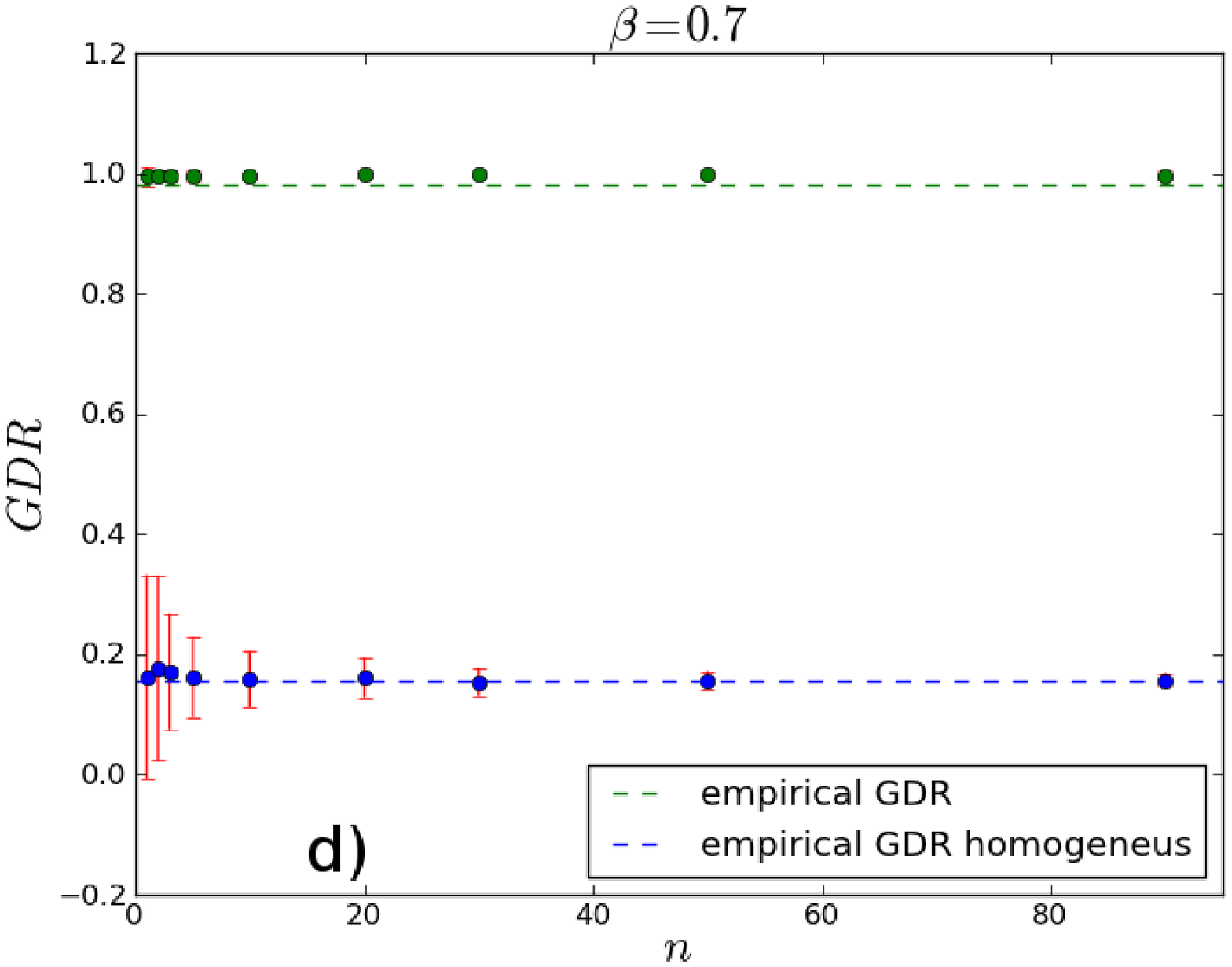} \\
  \multicolumn{2}{c}{\includegraphics[scale=0.3]{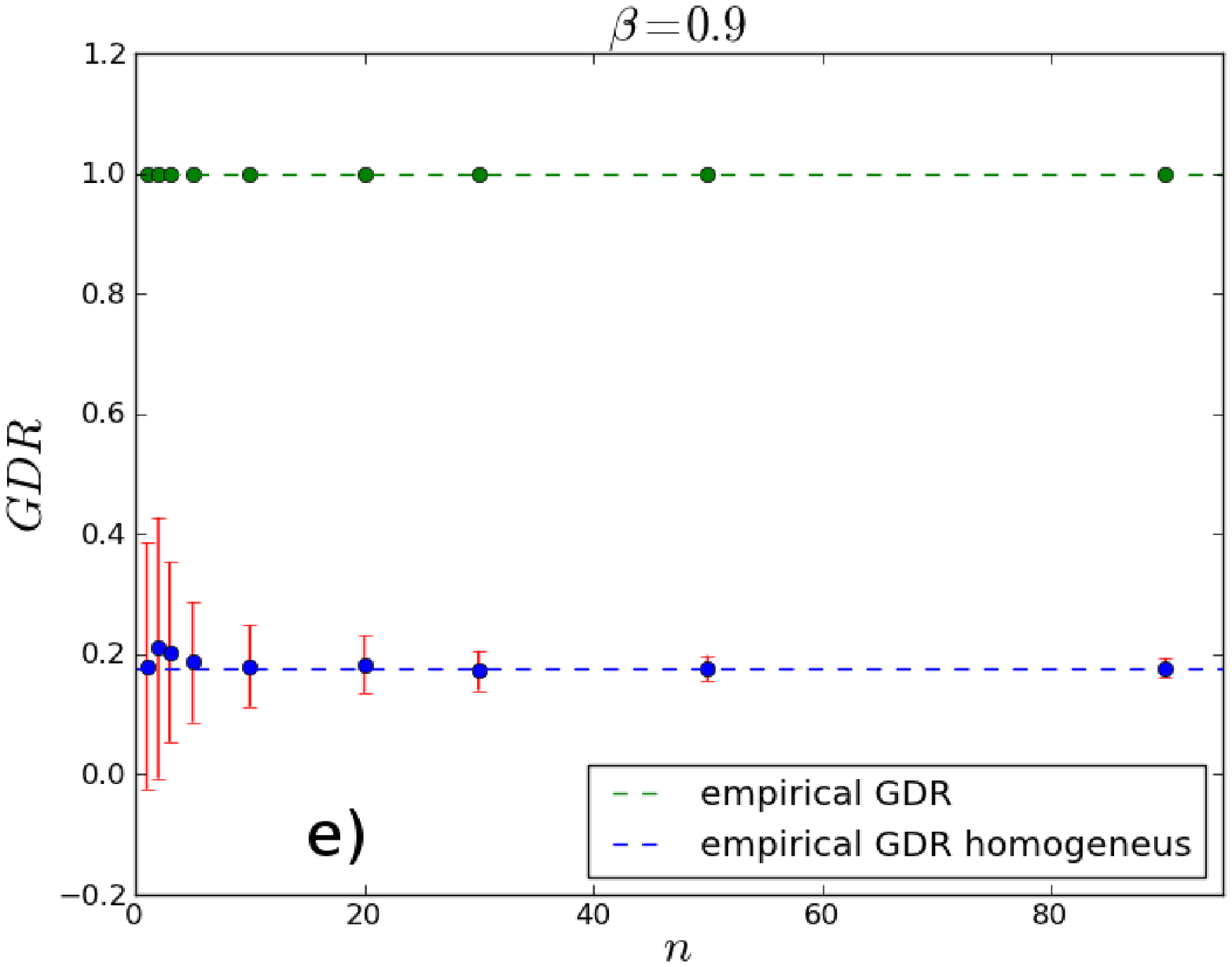} }

 \end{tabular}
\end{center}
 \caption{\label{fig:test_on_GDR}The pictures from top left represent respectively:  a) using
the impact rescaling factor $\alpha=0.1$ , compute the Group Debt Rank on the original WTW network with empirical weights (green dashed line), the average Group Debt Rank on the 100 bootstrapped
 networks with weights obtained using gravity model (green dots) and respective errors, the Group Debt Rank on the original WTW network with homogeneous weights (blue dashed line),
 and finally the average Group Debt Rank on the 100 bootstrapped
 networks with homogeneous weights (blue dots) and respective errors. b) same as in (a) but for $\alpha=0.3$ , c) same as in (a) but for $\alpha=0.5$, d)same as in (a) but for $\alpha=0.7$, e) same as in (a) but for $\alpha=0.9$} 
\end{figure}

We want to consider the impact in case a country defaults (it is not paying) while the actual adjacency matrix represents the economic value of the goods (the links are in the opposite direction). For this reason we transpose the WTW matrix and follow this procedure and we normalize imposing a row stochastic condition $\sum_{j}w_{ij} = 1$. The procedure for computing the GDR is the following (results on Fig. \ref{fig:test_on_GDR})

\begin{enumerate}
\item Compute the reference Group Debt Rank on the original WTW network with an initial shock $\phi = 0.1$ keep this value as reference (green dashed line in the plots)
\item Create a new network with the same topology of the WTW but with the weights replaced by the average weight of the WTW links (this value will be used in the simulation for all links)
\item Compute the GDR on a network with the same topology of the WTW and weight imposed homogeneous. Our goal is to study how close to this value (blue dashed line in the plot), using homogeneous values for the weights, the bootstrap method can go
\item  Bootstrap the networks with networks of size $n < N $ using homogeneous weights (computed as average of the weights of the only nodes that we start from in the simulation), compute the average GDR on 50 bootstrapped networks with homogeneous weights. Repeat the operation 100 times changing the starting set of nodes during the generation of the 50 networks, obtaining for each $n$ an average value with error (blue dots).

\item In the non homogeneous case (green dots) bootstrap the networks using weights according to a gravity model, where the weight of the link is the product of the GDPs of each node. To add an error, a "perturbation", on a such network we estimate empirically from the plot of $W_{ij}$ vs $GDP_i \cdot GDP_j$ the average variation of the weight $W_{ij}$ in function of the GDPs product. We then alter the corresponding adjacency matrix $W'_{ij}$ imposing, for each weight, a random normal error: $w'_{ij} = w_{ij} + \sigma N(0,1)$ where $\sigma$ is a standard deviation computed on $w_{ij}$ for the corresponding fixed value of $GDP_i \cdot GDP_j$. The new perturbed weight matrix is then transformed to maintain the row stochasticity.
\end{enumerate}  

\newpage

In Fig. \ref{fig:test_on_GDR} we plot the GDR for various $\alpha$ values, ranging from $0.1$ to $0.9$, in all cases the initial shock is $\psi = 0.1$, 10\% of the value of the trade of every country. From the pictures we can draw the following conclusions:
\begin{itemize}
\item There is a significant difference when using homogeneous weights or heterogeneous weights. Using a constant value for the weight of each link lead to underestimating the value of systemic risk as measured by  DebtRank
\item The reconstruction of the DebtRank values is already good for small subsets of nodes in the network.
\item Using a gravity model (even if simplified) improves the estimate of the GDR of the WTW network
\item The gap between the homogeneous GDR and the empirical one increases for larger values of the impact rescaling factor $\alpha$. This can be interpreted as follows: when the network effects are important the use of homogeneous weights in the dynamics leads to a larger error. Conversely when the network effects (reverberation) are less important the homogeneous weights are not so far from the \textit{true} value of the GDR.  
\end{itemize}

From this analysis we can conclude that the BM is good in reconstructing a non topological property such as the DebtRank but to achieve this goal one has to chose careful the weights of links because the use of an average value leads to inaccurate estimates, especially if the network effects are relevant. 

The impact of the single countries (DR) on the WTW network is shown in Tab. \ref{tab:drank} where the  impact rescaling factor $\alpha= 0.5$ and  the initial shock $\psi = 0.1$. As expected the biggest is the GDP the biggest is the corresponding DebtRank but with some variation due to network effects. Consider for instance a country like Canada, a big exporter of oil and minerals, its impact on the WTW will be larger than Germany that is a strong exporter of final goods. This analysis shows as the DebtRank measure is important to assess the distress (losses) propagation giving results that are not trivially expressed by the size of the countries.

\begin{table}[h]
\centering
\begin{tabular}{|l|l|l|}
\hline
Country & DebtRank & GDP rank(2000) \\
\hline
USA & 0.48 & 1 \\
JPN & 0.32 & 2 \\
CAN & 0.26 & 8 \\
CHN & 0.23 & 6 \\
DEU & 0.23 & 3\\
MEX & 0.18 & 10\\
GBR & 0.17 & 4 \\
FRA & 0.16 & 5 \\
ITA & 0.12 & 7 \\
NLD & 0.10 & 15 \\
KOR & 0.09 & 12 \\
TWN & 0.09 & 16\\
BEL & 0.08 & 20\\
ESP & 0.08 & 11\\
SGP & 0.07 & 39\\
MYS & 0.05 & 40\\
CHE & 0.05 & 18\\
BRA & 0.05 & 9 \\
IRL & 0.04 & 38\\
AUS & 0.04 & 14 \\
\hline
\end{tabular}
\caption{Table showing the DebtRank and the GDP rank (year 2000) for the 20 biggest countries in the WTW network. Notice that DebtRank is only in part respecting the same ranking given by the GDP of countries. Depending on the size of the export each country can be more or less affected by a shock on the default of the others countries. The values are computed using impact rescaling factor $\alpha = 0.5$ and $\psi = 0.1$, this numbers can interpreted in the following way: if the US will not pay the 10\% of their obligations to the rest of the world, the size of the trade is so big to cause a total loss of 48\% of the total WTW volumes. The amplification effect due to the network structure appears evident with a tool like DebtRank.}
\label{tab:drank}
\end{table}

\newpage
 
\section{Conclusions}\label{sec:conclusions}

In this paper we have proposed a new method to reconstruct the topology of a network using only partial information from its connections and an auxiliary non-topological property: the fitness associated to each node. This method is particularly useful to overcome the lack of topological information for several financial networks whose systemic risk must be measured.
Our approach allows to reconstruct the network using the topological information from one fraction of the nodes (i.e. their links) and a non topological property of each node derived from a fitness model. 

We tested the  network Bootstrap Method (BM) on the World Trade Web network, where we can use an accurate fitness model to describe its topology starting from the a non topological property (the GDP of the countries). We studied how well are reconstructed the following topological properties: the average density, the size, and the average degree of the main core. All these measures  are related to systemic risk for financial networks as  briefly presented in the introduction of this paper.

We found that the density of the links, the size of the main core, the average degree of the main core are reconstructed with an error varying from 1\% to 10\%, depending from the property examined, using with 5\% of the nodes (10 over the 185 nodes of the WTW network). An interesting finding is that the denser is a network the better is the reconstruction. The goodness of the reconstruction increases with the number of the nodes used as initial information and it is strongly dependent by the accuracy of the fitness model (that for the WTW is an accurate model describing how the links form across countries depending on their GDP and geographical distance).

The BM method was checked with another non topological property: the DebtRank a novel measure of systemic risk. We discovered that the method was really effective in evaluating this property also with a small number of starting nodes. We carried out a test using link weights derived from the gravity model of the WTW (the weight of a link is proportional to the product of the GDPs of each node) or using homogeneous (averaged) weights. 

In the case of homogeneous weights the BM estimates a value of Group Debt Rank (a measure of the DebtRank in a set of nodes) that is lower than the real one. This means that when the network is simulated using an average value for the weight there is a systematic bias in the evaluation of Systemic Risk measures such as the DebtRank. Conversely, in the case of non homogeneous weights, imposing more realistic values from the WTW fitness (gravity) model we obtain a more accurate estimation of the Group Debt Rank. This result stresses the importance, in the study of network systemic risk, to use a correct estimation of the weights and of the topological properties. Finally we notice that the bigger is the impact factor $\alpha$ in rescaling the nodes the greater is the distress propagation in the network captured by the Group Debt Rank.

We highlight that, for systemic risk, the network effects are responsible of an amplification of the distress. In fact the losses in the system due to a node failure (a default or a partial impossibility to pay) are bigger than the size of a country in terms of its ratio of GPD over the total market. In the paper we showed that countries that are not so big for GDP can have a significant impact on the WTW network depending on the size of their connections with others.

For what concerns possible future development, our work opens several challenges. 
To start with, we plan to test BM on other socio-economical networks, mainly financial ones. 
As written above, BM precision depends on how well fitness model describes real network. With WTW the fitness model works surprisingly well and it reproduces topological 
properties of any order \cite{squartini2011analytical}. We therefore need to test 
BM in more general cases, for example with networks where fitness model is 
less accurate and reproduces just some property.

\begin{acknowledgements}
We thank support from FET Open project FOC (255987) and Italian PNR project CRISIS-Lab
\end{acknowledgements}


\end{document}